\documentstyle[eqsecnum,floats,preprint,aps,prd,epsfig,rotate]{revtex}
\preprint{DAMTP R-99/108}
\date{16 August 1999}
\draft
\tighten
\begin{document}

\newcommand{\be}{\begin{equation}}
\newcommand{\ee}{\end{equation}}
\newcommand{\bea}{\begin{eqnarray}}
\newcommand{\eea}{\end{eqnarray}}

\title{Charged and rotating AdS black holes and their CFT duals}

\author{S.W. Hawking\thanks{S.W.Hawking@damtp.cam.ac.uk} and
H.S. Reall\thanks{H.S.Reall@damtp.cam.ac.uk}}

\address {\qquad \\ University of Cambridge \\ DAMTP\\
Silver Street\\
Cambridge, CB3 9EW \\United Kingdom}

\maketitle

\begin{abstract}

Black hole solutions that are asymptotic to $ AdS_5 \times S^5$ 
or $ AdS_4 \times S^7$ can
rotate in two different ways. If the internal sphere rotates then one
can obtain a Reissner-Nordstrom-AdS black hole. 
If the asymptotically AdS space rotates
then one can obtain a Kerr-AdS hole. One might expect
superradiant scattering to be possible in either of these
cases. Superradiant modes reflected off the potential barrier outside
the hole would be re-amplified at the horizon, and a classical instability
would result. We point out that the existence of a Killing vector 
field timelike everywhere outside the horizon prevents this from
occurring for black holes with negative action. Such black holes are
also thermodynamically stable in the grand canonical ensemble.
The CFT duals of these black holes correspond to a theory in an
Einstein universe with a chemical potential and a theory in a rotating
Einstein universe. We study these CFTs in the zero coupling limit. In
the first case, Bose-Einstein condensation occurs on the boundary
at a critical value of the chemical potential. However the
supergravity calculation demonstrates that this is not to be expected at
strong coupling.
In the second case, we investigate the limit in which the angular velocity of
the Einstein universe approaches the speed of light at finite
temperature. This is a new limit in which to compare the CFT at strong
and weak coupling. We find that the free CFT partition function and
supergravity action have the same type of divergence but the usual factor of
$4/3$ is modified at finite temperature.

\end{abstract}

\section{Introduction}

Black holes in
asymptotically flat space are often thought of as completely dead
classically. That is, they can absorb radiation and energy, but not
give any out. However, in 1969, Penrose
devised a classical process to extract energy from a rotating black
hole \cite{penrose}. 
This is possible because the horizon is rotating faster than
light with respect to the stationary frame at infinity. In other
words, the Killing vector $k$ that is time like at infinity is space like
on the horizon. The energy-momentum flux vector $J^{\mu} =
T^{\mu}_{\nu}k^{\nu}$ 
can therefore also be space like, even for matter
obeying the dominant energy condition. Thus the energy flow across the
future horizon of a rotating black hole can be negative: the Penrose
process extracts rotational
energy from the hole and slows its spin. This shows that 
rotating black holes are potentially unstable.

A nice way of extracting rotational energy
is to scatter a wave off the black hole \cite{zel,star}.  
Part of the incoming wave
will be absorbed, and will change the mass and 
angular momentum of the hole. By the first law of black hole mechanics
\be
 dM = \frac{\kappa}{8\pi} dA + \Omega dJ,
\ee
the changes in mass and angular momentum determine the change in area of
the horizon. The second law 
\be
 dA \ge 0
\ee
states that the area will increase in classical
scattering for fields that obey the dominant energy condition. For a
wave of frequency, $\omega$ and axial quantum number $m$ the
change of mass and the change of angular momentum  obey
\be
 \frac{dM}{dJ} = \frac{\omega}{m}.
\ee
The first and second second laws imply that the change of energy of
the black hole is negative when
\be
\label{eqn:supercond}
 \omega < m\Omega.
\ee
In other words, instead of part of the
incident wave being absorbed by the black hole and part reflected
back, the reflected wave would actually be stronger than the original
incoming wave. Such amplified scattering is called superradiance.

In a purely classical theory, a black hole is
won't lose angular momentum to massless fields like gravity. 
It is a different story, however, with
massive fields. A mass term $\mu$ for a scalar field, will prevent
waves of frequency $\omega < \mu$ from escaping to
infinity. Instead they will be reflected by a potential barrier at
large radius back into the hole. If they satisfy the condition for
superradiance then the waves will be amplified by scattering off the
hole. Each time the wave is reflected back, its amplitude will be
larger. Thus the wave will grow  exponentially and the black hole
will lose its angular momentum by a classical process.

One can understand this instability in the following way. In the WKB limit, a
mode with $\omega < \mu$ corresponds to a gravitationally bound
particle. If its orbital angular velocity is less than
the angular velocity of the black hole then angular momentum and energy
will flow from the hole to the particle. Orbits in
asymptotically flat space can have arbitrarily long periods so rotating
flat space black holes will always lose angular momentum to massive
fields, although in practice the rate is very low.

\bigskip

Charged fields scattering off an electrically charged black hole have similar
superradiant amplification \cite{den,dam,gary}. The condition 
is now 
\be
 \omega < q\Phi,
\ee 
where $q$ is the charge of the field and $\Phi$ the electrostatic 
potential difference between the horizon and infinity. 
There is, however, an important
difference from the rotating case. Black holes with regular horizons,
obey a Bogolomony bound, that their charges are not greater than their
masses, with equality only in the BPS extreme state. This bound
implies that the electro static repulsion between charged black holes,
can never be greater than their mutual gravitational attraction. 
The Bogolomony bound on the charge of a black hole implies
that $\Phi \le 1$ in asymptotically flat space. In a
Kaluza-Klein or super symmetric theory, the charges of fields will
generally obey the same BPS bound as black holes, with respect to
their rest mass i.e. $\mu \ge q$.  
This means the inequality for superradiance can
never be satisfied. One can think of this as a consequence of the
fact that the BPS bound implies that gauge repulsions never dominate
over gravitational attraction. It means that charged black holes in
supersymmetric and Kaluza Klein theories are classically stable. The
black hole can't lose charge by sending out a charged particle while
maintaining the area of the horizon, as it must in a classical
process.

So far we have been discussing superradiance and stability of 
black holes in asymptotically flat space. However, it should also be
interesting to study holes in anti-de Sitter space because the
AdS/CFT duality \cite{malda,gkp,witten} relates the properties of
these holes to thermal properties of a dual conformal field theory
living on the boundary of AdS \cite{witten,witten2}. A five
dimensional AdS analogue of the Kerr solution was constructed in
\cite{hhtr}. Reissner-Nordstrom-AdS (RNAdS) solutions of type IIB
supergravity were derived in \cite{andrew}\footnote{
More general charged black hole solutions of gauged
supergravity theories have also been discussed
in \cite{behrndt,duff} and their embedding in ten and eleven
dimensions was studied in \cite{embed}. 
Thermodynamic properties of charged AdS holes
have been discussed in \cite{andrew,braden,louko,peca,cvetic,cvetic2,andrew2}.
The thermodynamics of Kerr-Newman-AdS black holes in
four dimensions was recently discussed in \cite{cald}.}. 
These holes carry Kaluza-Klein charge coming from the
rotation of an internal $S^5$. The charged and rotating holes,
although appearing rather different in four or five dimensions, therefore
appear quite similar from the perspective of ten or eleven dimensional KK
theory: one rotates in the AdS space and the other in the
internal space. One aim of this paper is to
investigate whether these rotating black holes exhibit superradiance 
and instability and what that implies for the dual CFT. This CFT lives
on the conformal boundary of our black hole spacetimes, which is an
Einstein universe. 

The Kerr-AdS solution is discussed in section \ref{sec:kerrAdS}. We
find that a superradiant instability is possible only when the
Einstein universe on the boundary rotates faster than light. However
this can only occur when the black hole is suppressed in the
supergravity partition function relative to pure AdS. 

We discuss the RNAdS solutions in section \ref{sec:RNAdS}. We point
out that it is not possible for the internal $S^5$ to rotate faster
than light in these solutions and therefore superradiance cannot
occur, contrary to speculations made in \cite{andrew}. 
In particular this means that the extremal black holes, although 
not supersymmetric, are classically stable. It {\it is} possible for
the internal $S^5$ to rotate faster than light in $AdS_5 \times
S^5$. However such solutions have higher action than the corresponding
RNAdS solution (in a grand canonical ensemble) and are therefore
suppressed in the supergravity partition function and do not
affect the phase structure of the strongly coupled CFT.

A second aim of this paper is to compare the behaviour of the CFT at
strong and weak coupling. It was pointed out in \cite{gub1} that the
entropy of the strongly coupled theory (in flat space) is 
precisely $3/4$ that of the free theory - the surprise being that
there is no dependence on the t'Hooft parameter $\lambda$ or the
number of colours $N$. It has also
been noticed that the Casimir energy is the same for the free and 
strongly coupled theories \cite{bal}. This suggests that turning up
the temperature is similar to turning up the coupling.

We study the boundary CFT using a grand canonical ensemble. In the charged
case, this corresponds to turning on a chemical potential for a $U(1)$
subgroup of the $SO(6)$ R-symmetry group. In the rotating case, there
are chemical potentials constraining the CFT fields to rotate in the 
Einstein universe. In the free CFT, a $U(1)$ chemical potential
would cause Bose condensation at a critical value. This is not
apparent in the strongly coupled theory, which instead exhibits a
first order phase transition. Bose condensation
has been discussed in the context of
spinning branes \cite{gub3,kraus} but these discussions have referred
to CFTs in flat space, for which the energy of massless fields
starts at zero and Bose condensation would occur for any non-zero
chemical potential.

The rotating case was studied at high temperature in
\cite{hhtr,berman}. It was found that the factor of $3/4$ relating the
free and strongly coupled CFTs persists, even though there are extra
dimensionless parameters present that could have affected the result
\cite{berman}. At high temperature the finite radius of the spatial
sections of the Einstein universe is negligible so the theory behaves
as if it were in flat space. In the rotating case there is a new limit 
in which to study the behaviour of the CFT, namely the limit in 
which the angular velocity of the universe tends to the speed of
light. We find that the divergences in the partition functions 
of the free and strongly coupled CFTs are of the 
same form at finite temperature in this limit. 
We also examine how the $3/4$ factor is modified at finite temperature.

\section{Bulk and boundary rotation}

\label{sec:kerrAdS}

The three parameter Kerr-AdS solution in five dimensions 
was given in \cite{hhtr}. We shall start by reviewing the properties 
of this solution, which is
expected to be dual to the thermal properties of a strongly coupled
CFT in a rotating Einstein universe. 
We then investigate classical and thermodynamic stability.
Finally we calculate the partition function of the free CFT in
a rotating Einstein universe in order to compare the properties of the
strongly coupled and free theories.

\subsection{Five dimensional Kerr-AdS solution}

The five dimensional Kerr-AdS metric is \cite{hhtr}
\bea
 ds^2 &=& - \frac{\Delta}{\rho^2} (dt - \frac{a_1 \sin^2\theta}{\Xi_1}d\phi_1 -
 \frac{a_2 \cos^2\theta}{\Xi_2} d\phi_2)^2 +
 \frac{\Delta_{\theta}\sin^2\theta}{\rho^2} (a_1 dt -
 \frac{(r^2+a_1^2)}{\Xi_1} d\phi_1)^2 \nonumber \\
 && + \frac{\Delta_{\theta}\cos^2\theta}{\rho^2} (a_2 dt -
 \frac{(r^2+a_2^2)}{\Xi_2} d\phi_2)^2 + \frac{\rho^2}{\Delta} dr^2 +
 \frac{\rho^2}{\Delta_{\theta}} d\theta^2 \\
 && + \frac{(1+r^2)}{r^2 \rho^2}
 \left ( a_1a_2 dt - \frac{a_2 (r^2+a_1^2) \sin^2\theta}{\Xi_1}d\phi_1
 - \frac{a_1 (r^2 + a_2^2) \cos^2 \theta}{\Xi_2} d\phi_2 \right )^2, \nonumber
\eea
where we have scaled the AdS radius to one and 
\bea
\Delta &=& \frac{1}{r^2} (r^2 + a_1^2) (r^2 + a_2^2) (1 + r^2) - 2m;
\nonumber \\
\Delta_{\theta} &=& \left ( 1 - a_1^2 \cos^2\theta - a_2^2 
  \sin^2\theta \right ); \\
\rho^2 &=& \left ( r^2 + a_1^2 \cos^2\theta + a_2^2 \sin^2\theta \right);
\nonumber \\
\Xi_i &=& (1-a_i^2) \nonumber 
\eea
The metric is non-singular outside a horizon at $r=r_+$ provided
$a_i^2<1$. The angular velocities of the horizon in these
coordinates are
\be
 \Omega_i' = \frac{a_i (1-a_i^2)}{r_+^2 + a_i^2}
\ee
The corotating Killing vector field is
\be
 \chi = \frac{\partial}{\partial t} + \Omega_1'
 \frac{\partial}{\partial \phi_1} + \Omega_2' \frac{\partial}{\partial
 \phi_2},
\ee
which is tangent to the null geodesic generators of the horizon. 
These coordinates are not well-suited to demonstrating the
asymptotically AdS nature of this solution. A more appropriate set of
coordinates is defined as follows \cite{hhtr}
\bea
 T &=& t; \nonumber \\
\Xi_1 y^2 \sin^2\Theta &=& (r^2 + a_1^2) \sin^2\theta; \nonumber \\
\Xi_2 y^2 \cos^2\Theta &=& (r^2 + a_2^2) \cos^2\theta; \nonumber \\
\Phi_i &=& \phi_i + a_i t.
\eea
In these coordinates, the angular velocities become
\be
 \Omega_i = \frac{a_i (1+r_+^2)}{r_+^2+a_i^2}
\ee
and the corotating Killing vector field is
\be
 \chi = \frac{\partial}{\partial T} + \Omega_1
 \frac{\partial}{\partial \Phi_1} + \Omega_2 \frac{\partial}{\partial
 \Phi_2}.
\ee
The conformal boundary of the spacetime is an Einstein universe $R
\times S^3$ with
metric
\be
 ds^2 = -dT^2 + d\Theta^2 + \sin^2 \Theta d\Phi_1^2 + \cos^2
 \Theta d\Phi_2^2.
\ee
The action of the hole relative to an AdS background is calculated by
considering the Euclidean section of the hole obtained by analytically
continuing the time coordinate. To avoid a conical
singularity it is necessary to identify $(t,y,\Theta,\Phi_1,\Phi_2)$
with $(t+i\beta, y, \Theta, \Phi_1+i\beta\Omega_1,
\Phi_2+i\beta\Omega_2)$ where
\be
\label{eqn:beta}
 \beta = \frac{4 \pi (r_{+}^2 + a_1^2) (r_{+}^2 + a_2^2)}
 { r_{+}^2 \Delta'(r_{+})},
\ee
The same identifications must be made in the
AdS background in order to perform the matching. The action relative
to AdS is \cite{hhtr}
\be
 I = -\frac{\pi\beta (r_+^2+a_1^2)(r_+^2+a_2^2)(r_+^2-1)} {8G_5 r_+^2
(1-a_1^2)(1-a_2^2)},
\ee
where $G_5$ is Newton's constant in five dimensions.
The action is negative only for $r_+>1$. The boundary Einstein
universe inherits the above identifications from the bulk. 
The usual arguments \cite{gibbons} then show
that this identified Einstein universe is the appropriate 
background for path integrals defining thermal partition functions at
temperature
\be
\label{eqn:temp}
 T=\frac{1}{\beta} =  \frac{2r_+^6 + (1+a_1^2+a_2^2) r_+^4 - a_1^2
 a_2^2} {2\pi r_+ (r_+^2+a_1^2)(r_+^2+a_2^2)}
\ee
and
with chemical potentials $\Omega_i$ for the angular momenta $J_i$ of
matter fields in the Einstein universe. Matter is therefore
constrained to rotate in the Einstein universe; this is what is meant
by saying that the universe is rotating. The mass and angular momenta
of the black hole (using the coordinate $(T,\Phi_i)$) are \cite{hhtr}
\be
 M = \frac{3\pi m}{4(1-a_1^2)(1-a_2^2)}, \qquad J_i = \frac{\pi a_i m}
 {2(1-a_i^2)(1+r_+^2)}. 
\ee

\subsection{Stability of Kerr-AdS} 

In an asymptotically flat Kerr background there is a
unique (up to normalization) Killing vector field 
timelike near infinity i.e. $k=\partial /\partial t$. Near the horizon
there is an ergosphere - a region where $k$ becomes spacelike, and
energy extraction through superradiance becomes possible for modes
satisfying equation \ref{eqn:supercond}. In AdS,
superradiance would correspond to an instability of the hole. This is
because superradiant modes would be reflected back towards the hole by
a potential barrier (in the case of massive fields) or boundary
conditions at infinity (for massless fields) and reamplified at the
horizon before being scattered again. The hole would be classically
unstable and would lose angular momentum to a
cloud of particles orbiting it. The spectrum of fields in AdS is
discrete, and one might expect the threshold value of $\Omega$ for
superradiance to be given by the minimum of $\omega/|m|$ for fields
in the black hole background. However the presence of a black hole
changes the spectrum from discrete to continuous (since regularity at
the origin is no longer required) and it is not clear whether a positive
lower bound exists. Fortunately there is a simple argument that
demostrates the stability of Kerr-AdS for $|\Omega_i|<1$.

In Kerr-AdS, if $|\Omega_i|<1$ then the corotating Killing vector field
$\chi$ is timelike everywhere outside the horizon, so there
is a corotating frame that exists all the way out to infinity (in
contrast with the situation in flat space, where any rigidly rotating
frame, will necessarily move faster than light far from the axis of
rotation). The energy-momentum vector in this frame is $J^{\mu} =
T^{\mu}_{\nu} \chi^{\nu}$. If the matter obeys the dominant energy
condition \cite{he} then this is non-spacelike everywhere outside the
horizon. Let $\Sigma$ be a
spacelike hypersurface from the horizon to infinity with normal
$n_{\mu}$. The total energy of matter on $\Sigma$ is
\be
 E = -\int_{\Sigma} d^4 x \sqrt{h} n_{\mu} J^{\mu},
\ee
where $h$ is the determinant of the induced metric on $\Sigma$. The
integrand is everywhere non-positive so $E\ge 0$. The normal to the
horizon is $\chi_{\mu}$, so the energy flux density across the horizon
is $J^{\mu} \chi_{\mu}$, which is non-positive. If suitable boundary
conditions are imposed then energy will not enter the spacetime from
infinity. This means that if $E$ is evaluated on another surface
$\Sigma'$ lying to the future of $\Sigma$ then
\be
 E(\Sigma') \le E(\Sigma),
\ee
that is, $E$ is non-increasing function that is bounded below by
zero. Energy cannot be extracted from the hole: it is classically stable.

When $\Omega_i^2>1$, the corotating Killing vector field {\it does}
become spacelike in a region near infinity: this region rotates
faster than light. The above argument then
breaks down and an instability may occur. There are two different
limits in which $\Omega_i^2 \rightarrow 1$ \cite{hhtr}. The first is $a_i^2
\rightarrow 1$, which makes the metric become singular. 
The second is $r_+^2 \rightarrow a_i$ for 
which the metric remains regular. In
fact there is a range of $r_+^2 < a_i$ for which
$\Omega_i^2>1$. However since $a_i<1$, these black holes all have
$r_+<1$ and hence have positive action. They are therefore suppressed
relative to AdS in the supergravity partition function, so even if
these holes are unstable, the instability will not affect the phase 
structure of the CFT, although it may be of interest in its own
right.

\bigskip

We have demonstrated the absence of a classical instability when
$|\Omega_i|<1$. However we have not yet discussed thermodynamic
stability. In order to uniquely define the grand canonical ensemble, the
Legendre transformation from the extensive variables $(M,J_1,J_2)$ to
the intensive variables $(T,\Omega_1,\Omega_2)$ must be
single-valued. If this Legendre transformation becomes singular then
the grand-canonical ensemble becomes ill-defined. 
It is straightforward to calculate the determinant of the
jacobian:
\be
\label{eqn:jac}
 \det \frac{\partial (T,\Omega_1,\Omega_2)} {\partial (E, J_1,
J_2)} = \det \frac{\partial (T,\Omega_1,\Omega_2)} {\partial (r_+,a_1,a_2)} /
\det \frac{\partial (E, J_1, J_2)} {\partial (r_+,a_1,a_2)}.
\ee
The denominator vanishes if, and only if,
\bea
\label{eqn:denom}
 2(1-a_1^2 a_2^2) r_+^6 + (1 &+ & a_1^2 (2-a_1^2) +
 a_2^2(2-a_2^2) + a_1^2 a_2^2(3-a_1^2 a_2^2)) r_+^4 + {}\nonumber \\ 
 & & 2a_1^2 a_2^2
 (2+a_1^2 a_2^2) r_+^2 -a_1^2 a_2^2 (1-a_1^2 - a_2^2 - 3a_1^2 a_2^2)=0.
\eea
The right hand side can be written as
\be
 (1-a_1^2 a_2^2) \left[2r_+^6 + (1+a_1^2+a_2^2)r_+^4  - a_1^2
a_2^2\right] + \ldots
\ee
where the ellipsis denotes a group of terms that is easily seen to be
positive. The quantity in square brackets must also be positive in
order for a black hole solution to exist (as can be seen from equation
 \ref{eqn:temp}). Therefore equation
\ref{eqn:denom} cannot be satified. The numerator in equation
\ref{eqn:jac} vanishes if, and only if,
\be
  2r_+^6 - (1+a_1^2+a_2^2)r_+^4 +a_1^2 a_2^2 =0.
\ee
The right hand side is positive for $r_+>1$.
It has a negative minimum at a value of $r_+$ between
$0$ and $1$ and is positive at $r_+=0$ so there must be two roots
between $0$ and $1$. Let
$r_0$ denote the larger of these two roots. Black holes with $r_+>r_0$
are locally thermodynamically stable. However only those with $r_+>1$
have negative action, so the holes with $r_0<r_+<1$ are only metastable. 
The requirement of an invertible Legendre transformation therefore does not
affect the phase structure obtained from the action calculation.
Four dimensional
Kerr-AdS black holes behave in the same way \cite{cald}.

\subsection{Free CFT in a rotating Einstein universe}

\label{subsec:partition}

The high temperature limit of
free fields in a rotating Einstein universe was recently investigated in
\cite{hhtr,berman}. The usual factor of $4/3$ between the strongly
coupled and free CFTs was found to persist \cite{berman}. 
We wish to investigate a different limit, namely $\Omega_i
\rightarrow \pm 1$ at {\it finite} temperature. At finite
temperature, the finite size of the $S^3$ spatial sections of the
Einstein universe becomes significant. To compute the partition
function we need to know the spectrum of the CFT fields in the Einstein
universe. 

The Einstein universe has isometry group $R\times SO(4) = R\times
SU(2) \times SU(2)$, so we may classify representations of the
isometry group according to the Casimirs $(\omega,j_L,j_R)$ of $R$ and
the two $SU(2)$'s. The little group is $SO(3) = SU(2)/Z_2$. The
generators of this group are $J_i=J^{(L)}_i+J^{(R)}_i$, 
where $J^{(L)}_i$ and $J^{(R)}_i$ are
the generators of the two $SU(2)$ groups. Therefore the $SO(3)$
content of the representations of the isometry group is given by
angular momentum addition. The irreducible representation
$(\omega,j_L,j_R)$ will give a sum of irreducible representations of the
little group, with $|j_L-j_R| \le j \le j_L + j_R$. The lowest
eigenvalue $j=|j_L-j_R|$ is regarded as the spin. 
Therefore irreducible representations of the form $(\omega,j,j\pm s)$
describe particles of spin $s$. Parity invariance is obtained by
taking the direct sum $(\omega,j,j+s)+(\omega,j+s,j)$. These
representations may be promoted to representations of the conformal
group provided $\omega$ is suitably related to $j$ and $s$. The allowed values
of $\omega$ can be obtained by solving conformally invariant wave
equations on the Einstein universe. Alternatively they can be solved
on $AdS_4$, which is conformal to half of the Einstein
universe. This was done in \cite{breit}.
The scalar modes on $AdS_4$ can be extended to modes on the
Einstein universe. There are two different complete sets of modes on
$AdS_4$ however both sets are required for completeness on the Einstein
universe. The same happens for modes of higher spin.

The scalar modes form the representations $(j,j)$ of $SU(2)\times
SU(2)$. The energy eigenvalues are given by $\omega = J+1$ where
$J=2j$. The spin-$1/2$ modes form the representations
$(j,j+1/2)+(j+1/2,j)$ and have $\omega = J+1$ where $J=2j+1/2$. The
spin-1 modes form the representations $(j,j+1)+(j+1,j)$ with $\omega =
J+1$ and $J=2j+1$. In all cases the allowed values of $j$ are
$0,1/2,1,\ldots$. We have not taken account of the Casimir energy of
the fields because we have measured all energies relative to AdS
rather than using the boundary counterterm method \cite{bal} to calculate the
supergravity action.

The Killing vector fields of the Einstein universe form a 
representation of the Lie algebra of the
isometry group with $\partial/\partial\Phi_1 = J^{(L)}_3-J^{(R)}_3$ and
$\partial/\partial\Phi_2 = J^{(L)}_3+J^{(R)}_3$. Thus the quantum
numbers corresponding to rotations in the $\Phi_1$ and $\Phi_2$ directions
are $m_L-m_R$ and $m_L+m_R$ respectively.

We can now compute the partition functions for the CFT fields. In the
grand canonical ensemble, these are given by
\be
 \log Z = \mp \sum \log (1 \mp e^{-\beta (\omega - \Omega_1 (m_L-m_R) -
\Omega_2 (m_L+m_R))} ),
\ee
where the upper sign is for the bosons and the lower sign for the
fermions. Using the energy levels given above, the partition function
for a conformally coupled scalar field is given by
\bea
 \log Z_0 & = & -\sum_{J=0}^{\infty} \;
\sum_{m_L=-J/2}^{J/2} \; 
\sum_{m_L=-J/2}^{J/2}
 \log\left(1-e^{-\beta (J+1 - \Omega_1 (m_L - m_R) - \Omega_2
 (m_L + m_R))}\right) \nonumber \\
 & = & -\sum_{J=0}^{\infty} \; \sum_{m_L=-J/2}^{J/2} \; \sum_{m_R=-J/2}^{J/2}
 \log\left(1-e^{-\beta (J+1 - \Omega_+ m_L - \Omega_- m_R)}\right),
\eea
where $\Omega_{\pm} = \Omega_1 \pm \Omega_2$, the $J$-summation
runs over integer values and we have reversed the order of the $m_R$
summation. The partition function for a conformally coupled
spin-$1/2$ field is given by
\be
 \log Z_{1/2} = \sum_{J=1/2}^{\infty} \; \sum_{m_L = -(J+1/2)/2}^{(J+1/2)/2}
 \; \sum_{m_R = -(J-1/2)/2}^{(J-1/2)/2} \log \left(1+e^{-\beta (J+1 -
 \Omega_+ m_L - \Omega_- m_R)} \right) + (\Omega_+
 \leftrightarrow \Omega_-),
\ee
where the $J$-summation runs over half odd integer values. The first
term comes from the $(j+1/2,j)$ representations and the second from
the $(j,j+1/2)$ ones. The partition function for a conformally coupled
spin-$1$ field is given by
\be
 \log Z_1 = -\sum_{J=1}^{\infty} \; 
 \sum_{m_L = -(J+1)/2}^{(J+1)/2} \; \sum_{m_R
 = -(J-1)/2}^{(J-1)/2} \log \left(1-e^{-\beta (J+1 - \Omega_+
 m_L - \Omega_- m_R)}\right) + (\Omega_+ \leftrightarrow \Omega_-),
\ee
where the $J$-summation runs over integer values.

When $\beta$ is small, the sums in the above expressions may be
replaced by integrals. Doing so, one recovers the results of
\cite{berman}. For general $\beta$ we instead expand the logarithms as
power series. This gives
\be
 \log Z_0 =  \sum_{J=0}^{\infty} \sum_{m_L=-J/2}^{J/2}
 \sum_{m_R=-J/2}^{J/2} \sum_{n=1}^{\infty} \frac{1}{n} e^{-n\beta (J+1
 -\Omega_+ m_L - \Omega_- m_R)}.
\ee
We now interchange the orders of the $n$ and $J$
summations\footnote{This can be justified by cutting off the $J$
summation at $J=J_0$, proceeding as described in the text and
letting $J_0 \rightarrow \infty$ at the end.} The summations over
$m_L, m_R$ and $J$ can then be done (they are all geometric
series). One obtains
\be
 \log Z_0 = \sum_{n=1}^{\infty} \frac{e^{\beta n} \left(e^{2\beta
 n}-1\right)} {n\left(e^{\beta n (1-\Omega_1)}-1\right) \left(e^{\beta
 n (1+\Omega_1)}-1\right) \left(e^{\beta n (1-\Omega_2)}-1\right)
 \left(e^{\beta n (1+\Omega_2)}-1\right)}.
\ee
Similar calculations give
\be
 \log Z_{1/2} = \sum_{n=1}^{\infty} \frac{(-)^{n+1} 4 e^{3\beta n/2}
 \left(e^{\beta n}-1\right) \cosh (\beta n \Omega_1/2) \cosh (\beta n
 \Omega_2/2)}{n\left(e^{\beta n (1-\Omega_1)}-1\right) \left(e^{\beta
 n (1+\Omega_1)}-1\right) \left(e^{\beta n (1-\Omega_2)}-1\right)
 \left(e^{\beta n (1+\Omega_2)}-1\right)} 
\ee
and
\be
 \log Z_1 = \sum_{n=1}^{\infty} \frac{4\left(e^{\beta n}\cosh (\beta n
\Omega_1) -1 \right) \left( e^{\beta n}\cosh (\beta n \Omega_2) -1 \right)
+ 2\left(e^{2\beta n}-1\right)} {n\left( e^{\beta n
 (1-\Omega_1)}-1\right) \left( e^{\beta n (1+\Omega_1)}-1\right)
 \left( e^{\beta n (1-\Omega_2)}-1\right) \left( e^{\beta n
 (1+\Omega_2)}-1\right)}. 
\ee 
Note that all of these diverge as $\Omega_i \rightarrow \pm 1$. At first
sight this looks like Bose-Einstein condensation (since $\Omega_i$ is
a chemical potential) but this is misleading. The divergence does not
arise from the lowest bosonic energy level but from summing over
all of the modes (in particular the modes with largest $\Omega_+ m_L +
\Omega_- m_R$ for each $J$ \cite{hhtr}). Furthermore the fermion
partition function also diverges, so this is certainly not a purely
bosonic effect.

The particle content of the ${\cal N} = 4$ $U(N)$ super Yang-Mills
theory is $N^2$ gauge bosons, $4N^2$ Majorana fermions and $6N^2$
scalars. Adding the appropriate contributions from these fields, one
obtains the following asymptotic behaviour for the free CFT as
$\Omega_1 \rightarrow \pm 1$:
\be
 \log Z \approx \frac{2N^2}{\beta (1-\Omega_1^2)} \sum_{n=1}^{\infty}
\frac{(\cosh (\beta n \Omega_2/2) + (-)^{n+1})^2} {n^2 \sinh (\beta n
(1-\Omega_2)/2) \sinh (\beta n (1+\Omega_2)/2)},
\ee
and if we now let $\Omega_2 \rightarrow \pm 1$ then 
\be
 \log Z \approx \frac{8N^2}{\beta^2 (1-\Omega_1^2)(1-\Omega_2^2)}
\sum_{n=1}^{\infty} \frac{(\cosh (\beta n/2) + (-)^{n+1})^2} {n^3 \sinh
(\beta n)}.
\ee
The divergences as $\Omega_i \rightarrow 1$ are of the same form at
all temperatures. 
We are interested in comparing these divergences as for the free and 
strongly coupled CFTs. The partition function for the strongly coupled CFT is
given by the bulk supergravity partition function. For $r_+>1$ this is
dominated by the Kerr-AdS solution. To compare with the free CFT
results we introduce the stringy parameters. The five dimensional
Newton constant is related to the ten dimensional one by $1/G_5 =
\pi^3/G_{10}$, where the numerator is simply the volume of the internal
$S^5$. We are still using units for which the AdS length scale is
unity, which means that $\lambda^{1/4} l_s=1$ when we appeal to the
AdS/CFT correspondence. The ten-dimensional Newton constant is related
to the CFT parameters by $G_{10} = \pi^4/(2N^2)$ so $G_5 =
\pi/(2N^2)$. The supergravity action can then be written
\be
 I = -\frac{N^2\beta (r_+^2+a_1^2)(r_+^2+a_2^2)(r_+^2-1)} {4r_+^2
(1-a_1^2)(1-a_2^2)}.
\ee
Recall that in the bulk
theory there are two ways to take $\Omega_i \rightarrow 1$. However
one of these corresponds to a black hole suppressed relative to
AdS. We must therefore use the other limit, namely $a_i \rightarrow
1$. It is convenient to
use $r_+$ and $a_i$ instead of $\beta$ and $\Omega_i$ when comparing 
the partition functions for the strongly coupled and free CFTs. The
divergent factors in the free CFT are
\be
 \frac{1}{1-\Omega_i^2} = \frac{(r_+^2 + a_i^2)^2}{(r_+^4-a_i^2)(1-a_i^2)},
\ee
so both the strongly coupled and free CFTs have divergences
proportional to $(1-a_i^2)^{-1}$ in $\log Z$ as $a_i \rightarrow
1$. This generalizes the high temperature results of
\cite{hhtr,berman}.

\begin{figure}
\begin{picture}(0,0)(0,0)
\end{picture}
\centerline{\psfig{file=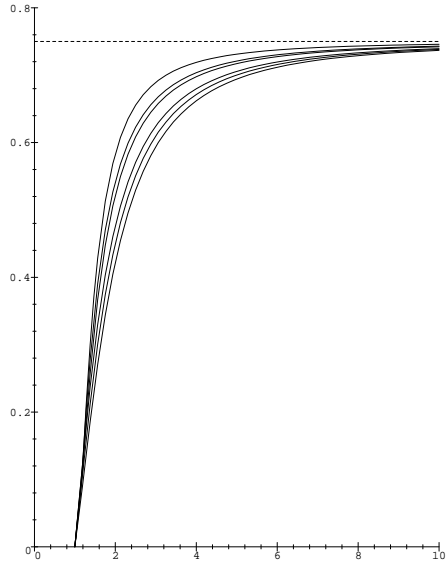,width=3in}}
\caption{Ratio of $\log Z$ for strongly coupled CFT to $\log Z$ for
free CFT as a function of $r_+$. From botton to top the curves are:
$a_1=a_2=0$; $a_1=0,a_2=0.5$; $a_1=0.5,a_2=0.5$;
$a_1 \rightarrow 1$, $a_2=0$;
$a_1 \rightarrow 1, a_2=0.5$; $a_1, a_2 \rightarrow
1$.}
\label{fig:ratio}
\end{figure}

The ratio
\be
 f(r_+,a_1,a_2) \equiv 
\frac{\log Z (\mathrm{strong})}{\log Z(\mathrm{free})} = -
\frac{I}{\log Z(\mathrm{free})},
\ee
is plotted as a function of $r_+$ for several cases of interest 
in figure \ref{fig:ratio}. 
At large $r_+$, $\beta \approx 0$, so the radius of the $S^3$ is
much larger than that of the $S^1$ of the Euclidean time
direction. The theory behaves as if it were in flat space. This is why
one recovers the flat space result \cite{gub1}
$f(\infty,0,0)=3/4$. The surprise pointed out in \cite{berman}
is that this is independent of $a_i$ i.e. $f(\infty,a_1,a_2)=3/4$. 
We have been studying a different limit, namely $a_i \rightarrow
1$. {\it A priori} there is no reason why this should commute with the
high temperature limit but it is straightforward to use the above
expressions to show that this is in fact the case, so all of the
curves in figure \ref{fig:ratio} approach $3/4$ at large $r_+$. 
At lower temperatures, there is still not much dependence of $f$ on
$a_i$. What is perhaps more surprising is how rapidly $f$ approaches
$3/4$: $f>0.7$ for $r_+=5.5$, which corresponds to $\beta \approx
0.58$ (for all $a_i$) so the radii of the time and spatial directions
are of the same order of magnitude and one might 
have expected finite size effects to be more important than they appear.

\subsection{The four dimensional case}

\label{sec:fourd}

The AdS/CFT correspondence relates the worldvolume theory of $N$
M2-branes in the large $N$ limit to eleven dimensional supergravity on $S^7$.
Four dimensional Kerr-AdS black holes are expected to be dual to the
worldvolume CFT in a rotating three dimensional Einstein
universe. For completeness we present the free CFT results for this case. 
The CFT is a free supersingleton field
theory \cite{nicolai}. There are eight real scalar fields and eight Majorana
spin-$1/2$ fields. The energy levels of these fields are $\omega =
j+1/2$ where $j=0,1,\ldots$ for the scalars and $j=1/2,3/2,\ldots$ for
the fermions \cite{bergshoeff}. 
The partition functions can be evaluated as above. For the scalars one obtains
\bea
 \log Z_0 &=& -\sum _{j=0}^{\infty} \sum_{m=-j}^{j} \log (1-e^{-\beta
(j+1/2-m\Omega)}) \nonumber \\
&=& \sum_{n=1}^{\infty} \frac{\cosh (\beta n/2)} {2n \sinh(\beta n
(1-\Omega)/2) \sinh (\beta n (1+\Omega)/2)},
\eea
and for the fermions,
\bea
 \log Z_{1/2} &=& \sum_{j=1/2}^{\infty} \sum_{m=-j}^{j} \log
(1+e^{-\beta (j+1/2-m\Omega)}) \nonumber \\
&=& \sum_{n=1}^{\infty} \frac{(-)^{n+1} \cosh (\beta n/2)} {2n \sinh(\beta n
(1-\Omega)/2) \sinh (\beta n (1+\Omega)/2)}.
\eea
Thus the partition function for the free CFT of an M2-brane is
\be
 \log Z = 8 \sum_{n\,\footnotesize{\mathrm{odd}}} \frac{\cosh (\beta n/2)} 
{n \sinh(\beta n (1-\Omega)/2) \sinh (\beta n (1+\Omega)/2)}.
\ee
At high temperature, one obtains
\be
 \log Z_0 \approx \frac{2 \zeta(3)}{\beta^2 (1-\Omega^2)}, \qquad
\log Z_{1/2} \approx \frac{3 \zeta(3)}{2\beta^2 (1-\Omega^2)}.
\ee
If $|\Omega| \rightarrow 1$ at finite temperature then
\be
 \log Z_0 \approx \frac{1}{\beta (1-\Omega^2)} \sum_{n=1}^{\infty}
\frac{1} {n^2 \sinh (\beta n/2)},
\ee
\be
 \log Z_{1/2} \approx \frac{1}{\beta
 (1-\Omega^2)} \sum_{n=1}^{\infty} \frac{(-)^{n+1}} {n^2 \sinh(\beta n/2)},
\ee
and 
\be
 \log Z \approx \frac{16}{\beta(1-\Omega^2)} \sum_{n\,
\footnotesize{\mathrm{ odd}}} \frac{1}{n^2 \sinh (\beta n/2)}.
\ee 
The divergence is of the same form as that obtained from the bulk
supergravity action in the limit $|a| \rightarrow 1$ \cite{hhtr}.

\section{Bulk charge and boundary chemical potential}

\label{sec:RNAdS}

It was shown in \cite{andrew} how to obtain Einstein-Maxwell theory
with a negative cosmological constant from KK reduction of IIB 
supergravity on $S^5$. The reduction ansatz for the metric
is\footnote{We have rescaled the electromagnetic potential relative to
that of \cite{andrew}.}
\be
 ds^2=g_{\mu\nu}dx^{\mu}dx^{\nu}  + \sum_{i=1}^3 [d\mu_i^2 + 
\mu_i^2(d\phi_i + A_{\mu}dx^{\mu})^2],
\ee
where $g_{\mu\nu}$ is a five dimensional metric, $\mu_i$ are direction
cosines on the $S^5$ (so $\sum_{i=1}^3 \mu_i^2=1$) and the $\phi_i$
are rotation angles on $S^5$ in three orthogonal planes (when embedded
in $R^6$). Non-vanishing $A_{\mu}$ corresponds to rotating the $S^5$
by equal amounts in each of these three planes, and gives a Maxwell
electromagnetic potential in five dimensions after KK reduction. 
The ansatz for the Ramond-Ramond 5-form is given in \cite{andrew}. 

\subsection{$AdS_5\times S^5$ with electrostatic potential}

The simplest solution of the Einstein-Maxwell system with negative
cosmological constant is $AdS_5$ with metric
\be
 ds^2 = -U(r)dt^2+U(r)^{-1}dr^2+r^2 d\Omega_3^2
\ee
where
\be
 U(r)= 1+r^2
\ee
and a constant electrostatic
potential $A=-\Phi dt$ with $\Phi=$const. Increasing $\Phi$
corresponds to increasing the angular velocity of the internal
$S^5$. A point at fixed $\mu_i$ and $\phi_i$ on the $S_5$ moves on an
orbit of $k=\partial/\partial t$. This has norm
\be
 k^2 = -U(r) + \Phi^2,
\ee
so $k$ will be spacelike in a region near $r=0$ when $\Phi^2>1$. 
This means that the $S^5$ rotates faster than light
near the origin in $AdS_5$ when $\Phi$ is large. The $t$-direction
becomes spacelike and an internal direction becomes timelike,
indicating an instability. If a BPS particle were added to this
solution in the grand ensemble then, near the origin, its negative
electric potential energy would exceed its rest mass, so the most probable
configuration would involve an infinite number of particles. 

In the AdS/CFT correspondence, a bulk electromagnetic potential $A$
couples to a conserved current of the boundary theory
\cite{gkp,witten}. In our case, the electromagnetic potential
is associated with the $U(1)$ obtained by taking equal charges for the
three $U(1)$ groups in the $U(1)^3$ Cartan subalgebra of the $SO(6)$
KK gauge group. The CFT current is therefore obtained by taking the
same $U(1)$ subgroup of the $U(1)^3$ Cartan subalgebra of the $SU(4)$
R-symmetry group of the boundary CFT. The coupling of the bulk gauge
field to the boundary current is $- A_i j^i$, where 
\bea
 j_i &=&  r^2 \sum_{k=1}^3 \mu_k^2 \partial_i \phi_k +
\mathrm{fermions} \nonumber \\
 &=&  \sum_{k=1}^3 \left(X^{2k-1} \partial_i X^{2k} 
 - X^{2k} \partial_i X^{2k-1}\right) + \mathrm{fermions}
\eea
where $X^k$ are the usual scalar fields of the ${\cal N}=4$ $SU(N)$ super
Yang-Mills theory and there is a suppressed sum over $N$.
The fermionic
contribution should be straightforward to calculate although we
shall not do so. 

Taking $A=-\Phi dt$ corresponds to turning on a chemical potential
$\Phi$ for the $U(1)$ charge 
in the boundary theory. In the free CFT,
Bose-Einstein condensation will result when this chemical potential
equals the lowest bosonic energy level, which is $\omega =
1$ (see section \ref{subsec:partition}). Thus BE condensation 
occurs in the free CFT when 
$\Phi = \pm 1$ (the two signs refer to particles and
anti-particles respectively). This is precisely the critical
value of $\Phi$ for which the internal sphere rotates at the speed of
light. 

\subsection{Reissner-Nordstrom-AdS black holes}

Solutions of type IIB supergravity describing Reissner-Nordstrom-AdS
black holes with an internal $S^5$ were given in \cite{andrew}.
The five dimensional metric can be written
\be
 ds^2 = -V(r)dt^2 + V(r)^{-1}dr^2 + r^2 d\Omega_3^2,
\ee
with
\bea
 V(r) & = & 1-\frac{M}{r^2}+\frac{Q^2}{r^4}+r^2 \nonumber \\
 & = &\left(1-\frac{r_+^2}{r^2}\right)
 \left(1-\frac{r_-^2}{r^2}\right)
 \left(1+r^2+r_+^2+r_-^2 \right),
\eea
where $M$ and $Q$ measure the black hole's mass and charge
and $r_{\pm}$ are the outer and inner horizon radii. The electromagnetic
potential in a gauge regular on the outer horizon is
\be
 A = (\Phi(r_+)-\Phi(r)) dt.
\ee
where
\be
 \Phi(r) = \frac{Q}{r^2}.
\ee
Once again we can compute the norm of the Killing vector
field $k=\partial/\partial t$ with respect to the 10-dimensional
metric. This is
\bea
 k^2 & = & -V(r) + \frac{Q^2}{r_+^4} \left(1-\frac{r_+^2}{r^2}\right)^2
\nonumber \\
 & = & - \left(1-\frac{r_+^2}{r^2}\right) \left[r^2
\left(1-\frac{r_-^2}{r^2}\right) + \left(1-\frac{r_-^2}{r_+^2}\right)
\left(1 + r_+^2+r_-^2\right) \right]
\eea
and this is negative for $r>r_+$. 
Hence the internal $S^5$ never rotates faster
than light outside the black hole: there is an everywhere timelike
Killing vector field outside the hole. The stability argument we used
for Kerr-AdS can be therefore also be applied in this case to
conclude that energy extraction from RNAdS black holes is impossible.

\begin{figure}
\begin{picture}(0,0)(0,0)
\put(310,-120){$T$}
\put(172,3){$\Phi$}
\put(170,-50){$1$}
\end{picture}
\centerline{\psfig{file=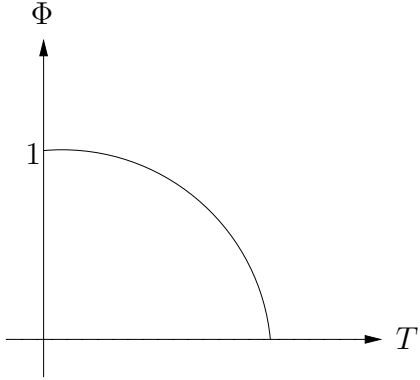,width=2.in}}
\caption{Phase diagram for Reissner-Nordstrom-AdS. AdS is preferred 
in the region near the origin.}
\label{fig:phase}
\end{figure}

The action $I$ of the black hole relative to AdS is \cite{andrew}
\be
 I = \frac{\pi}{8 G_5}\beta (r_+^2 (1-\Phi(r_+)^2) - r_+^4)
\ee
where the inverse temperature is
\be
 \beta = \frac{2\pi r_+}{1-\Phi(r_+)^2 + 2r_+^2}.
\ee
This action can be related to the thermal partition function
of the strongly coupled gauge theory on the boundary
\cite{witten,witten2}. We are interested in the partition function in
the grand canonical ensemble, for which the chemical potential and
temperature are fixed on the boundary. The phase diagram was given
in \cite{andrew} and reproduced in figure \ref{fig:phase}. 
There is a
region near the origin of the $\Phi-T$ plane for which $I$ is
positive so AdS is preferred over the black hole. Everywhere else, $I$
is negative so the hole is preferred. A first order phase transition
occurs when $I$ changes sign. Note that the internal sphere
does not reach the speed of light anywhere in this diagram. The closest
one can get is to let $T$ tend to zero whilst increasing $\Phi$ to the
critical value in AdS. As soon as the critical value is reached,
an extreme black hole of vanishing horizon radius becomes preferred over pure
AdS. Thermodynamic stability of RNAdS was discussed in
\cite{andrew2}. In the grand canonical ensemble, it was found that
black holes with positive action are stable. 

This phase diagram is very different from that of the free boundary
CFT, which only has a phase transition at the critical value of
$\Phi$. The strongly coupled CFT does not exhibit a phase transition as
the chemical potential is increased 
at high temperature, unlike the free CFT. Thus at finite
chemical potential, the thermal partition functions of
the free and strongly coupled CFTs in an Einstein universe differ by
much more than a simple numerical factor, even at high temperature.

In four dimensions the situation is identical. The lowest bosonic
energy level is $\omega = 1/2$ (see section \ref{sec:fourd}) so
Bose condensation in the free field theory
occurs at $\Phi=1/2$, which is the critical value for the internal
sphere in $AdS_4 \times S^7$ to rotate at the speed of light (the KK
ansatz for the four dimensional case was given in \cite{andrew}). The
phase structure of the strongly coupled theory is qualitatively
identical the the one in figure \ref{fig:phase} except that the phase
transition occurs at $\Phi=1/2$ on the $T=0$ axis.

\section{Discussion}

We have studied the stability of rotating asymptotically
$AdS_5 \times S^5$ and $Ads_4\times S^7$ solutions of supergravity. A classical
instability can occur if either the boundary of the $AdS$ space 
or the internal $S^5$ rotates faster than light. 
However this occurs only when the solution
has positive action relative to $AdS$ and is therefore suppressed in
the supergravity partition function.   
Reissner-Nordstrom-AdS solutions do not exhibit a
superradiant instability but small Kerr-AdS solutions may do, although
a proof would involve studying wave equations in
Kerr-AdS. We have also studied 
quantum local thermodynamic stability and found that
the solutions that are not locally stable have positive action. AdS
space is preferred in the domain of the black hole parameters for
which the holes are locally unstable. This is to be contrasted with
the charged black holes of \cite{cvetic}, for which there was a region
of parameter space in the grand canonical ensemble where the black
holes were preferred over AdS but not locally stable.

We have compared the strongly coupled coupled and free boundary CFTs
in an Einstein universe. 
When the Einstein universe rotates, we find that the free and strongly
coupled theories have the same type of divergence as the angular
velocities approach the speed of light at finite temperature. 
The factor of $3/4$ relating the
partition functions is recovered at high temperature in the Einstein
universe since then the radius of curvature of the $S^3$ spatial
sections is negligible compared with the radius of curvature of the
Euclidean time direction. That this factor is independent of the
angular velocities at high temperature was noticed in \cite{berman}; 
we have found that it does not vary greatly with angular velocity at
lower temperatures either.

Free field theory in the Einstein universe is not a good guide to the
properties of the strongly coupled theory at finite $U(1)$ chemical potential
since the former would undergo Bose condensation at a critical
chemical potential whereas the latter does not. 
Studying this in the
Einstein universe allows us to avoid the problems associated with
chemical potentials in CFTs in flat space. The critical chemical
potential at which Bose condensation occurs is the value of the
potential for which the internal sphere in $AdS_5 \times S^5$ rotates
at the speed of light. However the phase transition in the strongly
coupled theory only occurs at this value in the limit of zero
temperature.

\bigskip

\centerline{{\bf Acknowledgements}}

HSR would like to thank Andrew Chamblin, Roberto Emparan, Clifford
Johnson and Marika Taylor-Robinson for helpful discussions.

\end{document}